# Effect of Rare-Earth Ions on an Electric Polarization Induced by the Phase Separation Domains in RMn$_2$O$_5$ (R = Er, Tb)

B. Kh. Khannanov[1], E. I. Golovenchits[1], and V. A. Sanina[1],

[1.] Ioffe Institute, St. Petersburg, 194021 Russia

**Abstract**—The effect of rare-earth ions (R = Er$^{3+}$, Tb$^{3+}$) with strong spin-orbit coupling on the dielectric properties and the electric polarization induced by local polar phase separation domains in RMn$_2$O$_5$ multiferroics has been studied. These parameters were found to be qualitatively distinguished from those studied before in GdMn$_2$O$_5$, in which Gd$^{3+}$ ion in the ground $^8S_{7/2}$ state is weakly bounded with the lattice. It is shown that the properties of the polar phase separation domains, which form in the subsystem of Mn$^{3+}$ and Mn$^{4+}$ ions, are substantially dependent on the values of crystal fields, in which these domains exist.

1. INTRODUCTION

Manganites RMn$_2$O$_5$ (R is a rare-earth ion, Bi) are typical representatives of II-type multiferroics, in which the ferroelectric ordering is induced and controlled by a magnetic order. The characteristic values of the Curie $T_C$ and the Neel $T_N$ temperatures are 30–35 K and 40–45 K, respectively [1, 2]. Until recently it was assumed as an established fact that RMn$_2$O$_5$ are characterized by centrosymmetric space group *Pbam* that forbids the existence of electric polarization. To explain the polarization observed in RMn$_2$O$_5$ at $T < 30$–35 K, the exchange striction model was proposed. Exchange striction appears when considering a charge ordering of pairs of the neighboring manganese ions of different valences (Mn$^{3+}$ and Mn$^{4+}$) along axis *b* with alternating ferromagnetic and antiferromagnetic orientations of spins of these ions. A significant excess of the ferromagnetic interaction (double exchange) over the antiferromagnetic interaction (indirect exchange) leads to the appearance of the exchange striction that violates the centrosymmetricity of a lattice and the appearance of the ferroelectric ordering along axis *b* [3]. Thus, it was assumed that the electric polarization in $RMn_2O_5$ exists only at temperatures $T \leq T_C$.

In [4], it was reported on the structural studies of a series of $RMn_2O_5$ crystals with various R ions at room temperature using the resonant synchrotron X-ray diffraction and the geometric optimization based on the first-principles calculations. Those authors revealed, in addition to intense reflections corresponding to space group *Pbam*, significantly weaker reflections that could not be described by space group with the central symmetry. Using physical arguments, they stated that the real structure of $RMn_2O_5$ is described by monoclinic noncentrosymmetrical space group *Pm* that allows the existence of the polarization along axis *b*. In [4], it was concluded that a ferroelectric ordering of other nature (unlike exchange striction) must exist in $RMn_2O_5$ up to room temperature. At $T < 30–35$ K, two of these mechanisms coexist. The nature of the high-temperature ordering was not discussed in [4].

In some $RMn_2O_5$ (R = Eu, Gd, Bi), we studied the temperature evolution of the dielectric properties (permittivity and conductivity) over the wide temperature range 5–350 K [5–9]. Two facts were established. First, the free dispersion maxima of permittivity and dielectric losses, which are characteristic of the phase transition to the ordered ferroelectric state, were observed along axis *b* only near $T = T_C = 30–35$ K. In the temperature range 35–350 K, no such maxima were observed along all the axes. This fact shows that there is no a second high-temperature phase transition to the ordered ferroelectric state in this temperature range. Second, in the paraelectric state up to 350 K, we observed the anomalies of the permittivity and the conductivity depending on frequency and characteristic of the local polar domains. The splitting of the Bragg peaks into two reflections [5–9] showed that the sizes of these domains are sufficient for a different structural ordering, unlike that in the matrix, to appear in them.

A characteristic feature of $RMn_2O_5$ is the existence of the same numbers of manganese ions $Mn^{3+}$ (containing three $t_{2g}$ and one $e_g$ electrons on the $3d$ shell) and $Mn^{4+}$ (with three $t_{2g}$ electrons on the $3d$ shell), which determines the conditions for the formation of a dielectric charge ordering. The $Mn^{4+}$ ions have the octahedral oxygen environment and are located in layers with $z = 0.25c$ and $(1 - z) = 0.75c$. The $Mn^{3+}$ ions have an eccentric local environment in the form of pentagonal pyramids and are disposed in layers with $z = 0.5c$. Ions $R^{3+}$ with the environment similar to that of $Mn^{3+}$ ions are disposed in layers with $z = 0$ [10]. The charge ordering and a finite probability of the transfer of $e_g$ electrons between the $Mn^{3+}$–$Mn^{4+}$ ion pairs (double exchange [11, 12]) are key factors which determine the electric polar states of $RMn_2O_5$ at all temperatures. As noted above, the low-temperature ferroelectric state at $T \leq T_C$ is due to the charge ordering along axis $b$ [3]. On the other hand, the transfer of $e_g$ electrons between the $Mn^{3+}$–$Mn^{4+}$ ion pairs arranged in the neighboring layers perpendicular to axis $c$ leads to the formation of local phase separation domains with another distribution of $Mn^{3+}$ and $Mn^{4+}$ ions as compared to the initial crystal matrix. As is shown in [5–9], these local domains in $RMn_2O_5$ are polar and exist from the lowest temperatures to temperatures higher than room temperature.

The local polar domains containing $Mn^{3+}$ and $Mn^{4+}$ ions form in $RMn_2O_5$ due to the phase-separation processes similar to the case in $LaAMnO_3$ (A = Sr, Ca, Ba) [12, 13]. The restricted phase separation domains form in the initial crystal matrix due to the self-organization processes determined by a finite tunneling probability of $e_g$ electrons between the ($Mn^{3+}$–$Mn^{4+}$) ion pairs and exist from the lowest temperatures to temperatures higher than room temperature. At low temperatures ($T < T_C \approx 30$–$40$ K), the local phase separation domains in $RMn_2O_5$ are isolated 1D superlattices of ferromagnetic layers containing $Mn^{3+}$ and Mn4+ ions in various proportions. In them, a set of ferromagnetic resonances [14–16] and the electric polarization [7–9] were observed in the directions of the magnetic and electric fields, respectively. At temperatures higher than 180 K, the interaction appears between local polar domains isolated before, and it forms 2D superstructures perpendicular to axis *c*. In these superstructures, the initial matrix layers alternated to the phase separation domains. At room temperature, the layer widths were 700–900 Å [5, 6].

When studying the electric polarization induced by local polar domains in $RMn_2O_5$ (R = Gd, Bi) [7–9], it was shown that, in these multiferroics, the local polar domains are the phase separation domains and they have the local ferroelectric ordering. The polarity of the phase separation domains in $RMn_2O_5$ is due to the following two factors. In the phase separation domains, the double exchange caused by the transfer of $e_g$ electrons between the $Mn^{3+}$–$Mn^{4+}$ ion pairs leads to the fact that Yahn–Teller $Mn^{3+}$ ions occupy positions of $Mn^{4+}$ ions (oxygen octahedra) and deform these octahedra. In turn, $Mn^{4+}$ ions (having lower sizes compared to $Mn^{3+}$ ions) are in eccentric pentagonal pyramids and also additionally distort them. Both these factors lead to the noncentrosymmetricity of the phase separation domains and to their polarity. These domains form the superparaelectric state that is in a frozen superparaelectric states below of some temperatures. The response of local polar domains in this state to the applied electric field has a form of hysteresis loops of the electric polarization with the residual polarization. As shown in [7–9], the frozen superparaelectric state exists to the temperatures, at which the kinetic energy of free charge carriers is compared with the barrier heights at the domain boundaries.

The frozen superparaelectric state for local ferroelectric domains in a dielectric centrosymmetric matrix was considered theoretically in [17], but it was observed experimentally in $RMn_2O_5$ (R = Gd, Bi) for the first time in [7–9]. In this case, it was shown that local polar phase separation domains in RMn$_2$O$_5$ are multiferroics, since they are controlled external electric and magnetic fields and exist to temperatures which are higher than $T_C$ of the low-temperature ferroelectric ordering. In those works, it was assumed that the low-intensity eccentric reflections in the structural study at room temperature in [4] correspond to such type of the local polar domains.

The main problem of this work is the experimental observation and the study of the properties of the electric polarization induced by phase separation domains in RMn$_2$O$_5$ (R = Er and Tb), and the comparison of their properties with the properties of GdMn$_2$O$_5$ studied before [7–9]. A set of crystals was chosen by the following reasons. Ion Gd3+ (ground state $^8S_{7/2}$) has the maximum spin among R-ions and it is weakly related to the lattice in the ground $S$ state, since the spin–orbit interaction is absent in this ion. The ground state of Tb$^{3+}$ ($^7F_6$, $S = 3$, $L = 3$) is characterized by a high magnetic moment ($J = 9.7\mu_B$), to which both spin and orbital moments ($S = 3$, $L = 3$, respectively) contribute, and there is spin–orbit coupling. Ions Tb$^{3+}$ are described in the extreme strong anisotropic Ising approximation that rigidly fixes the orientation of their moments in plane *ab* [18]. The ground state of Er$^{3+}$ ion ($^4I_{15/2}$, $S = 3/2$, $L = 6$) also has a high magnetic moment ($J = 9.6\mu_B$), to which the orbital moment make the main contribution and there is a strong crystal field that rigidly orientates the Er ion moments along axis *c* of strong single-ion anisotropy [18]. The magnetic moments of R ions are bounded by the exchange interaction with Mn ions. In addition, R ions strongly change the crystal field in which the Mn ion subsystem exists. It is interesting to compare the influences of crystal fields on the properties of the Mn subsystem when taken into account the spin-orbit interaction of R$^{3+}$ ions.

## 2. EXPERIMENTAL

The RMn$_2$O$_5$ (R = Er, Tb) single crystals were grown by the spontaneous crystallization from solution–melt [19, 20]. They were 2–3-mm-thick plates with area 3–5 mm$^2$.

To measure the dielectric properties and the polarization, we fabricated flat capacitors 0.3–0.6 mm in thickness and with area of 3–4 mm$^2$. The permittivity and the conductivity were measured using a Good Will LCR-819 impedance meter in the frequency range 0.5–50 kHz in the temperature range 5–330 K. The electric polarization was measured by the PUND (Positive Up Negative Down) method [21–23]. We used the PUND method adapted to measurements of the polarization of local polar domains with a local conductivity and described in [7–9]. The PUND method enables one to correctly take into account the contribution of the conductivity to the measured electric polarization loop. In this method, the response of internal polarization *P* can be separated when applying a series of positive *P*1–*P*2 and negative *N*1–*N*2 voltage pulses to the sample. In this case, we recorded the independent (*P*1–*P*2) and (*N*1–*N*2) curves of the response of the effective changes in the polarization on this series of voltage pulses. The PUND method is based on the difference of the dynamics of the responses of the internal polarization and the conductivity on the pulses of the applied electric field *E*. As a pulse is turned off, the polarization relaxes significantly slower than the conductivity does. The time intervals between (*P*1–*P*2) and (*N*1–*N*2) pulses must be such that the internal polarization was slightly changed for this time, whereas the relaxation of the conductivity was finished completely. In common bulk ferroelectrics with a domain structure, such intervals can be several seconds. In the case of the response from dynamic polar domains of the phase separation, which quite quickly reestablishes the equilibrium state after switching off field *E*, it is necessary to decrease the intervals between (*P*1–*P*2) and (*N*1–*N*2) pulses. However, in this case, it is necessary, as before, to provide the fulfillment of the condition that the internal polarization was slightly changed for this time, whereas the relaxation of the conductivity was finished completely. The real internal polarization is calculated by the subtraction of pair (*P1–P2*) and (*N1–N2*) pulses from one another.

## 3. RESULTS AND DISCUSSION

*3.1. Dielectric Properties and the Electric Polarization in ErMn2O5*

Figures 1a‑1f show the temperature dependences of permittivity ε' for a number of frequencies along axes *a*, *b*, *c* (Figs.1a, 1c, and 1e, respectively) and the conductivity σ along these axes (Figs. 1b, 1d, and 1f, respectively). The insets in the plots for the conductivity show the temperature dependences of the relative local conductivity $\sigma_{loc} = (\sigma_{AC} - \sigma_{DC})/\sigma_{DC}$ for the corresponding axes. According to the accepted picture of the dispersion of the conductivity of inhomogeneous media which contain local nanodomains in a homogeneous matrix [24], the dispersion free low-frequency conductivity characterizes a percolation conductivity of the matrix $\sigma_{DC}$. On the other hand, the frequency dependent conductivity $\sigma_{AC}$ (the higher the frequency, the higher is the conductivity) characterizes the local conductivity of individual domains. It is convenient to characterize the ratio of the local and percolation conductivities by the local conductivity $\sigma_{loc} = (\sigma_{AC} - \sigma_{DC})/\sigma_{DC}$ [24].

We are dealing with the real part of the conductivity $\sigma_1 = \omega\varepsilon''\varepsilon_0$ [24] that is calculated from dielectric losses ε'' (we measure dielectric loss tangent tanδ = ε''/ε'). Here, ω is the angular frequency and $\varepsilon_0$ is the permittivity of free space. Conductivity $\sigma_1$ denoted as σ in what follows is dependent on the frequency and also on temperature. We assume that, in our case, the local conductivity is due to the phase separation domains, and the percolation conductivity corresponds to the initial $ErMn_2O_5$ matrix.
The dispersion free maximum of the permittivity that demonstrates the low-temperature ferroelectric phase transition (the inset in Fig. 1c) is observed only along axis *b* near $T_C$ = 37 K. Along all the directions, in the temperature range 5‑150 K, permittivity ε' = 20 independent of temperature is observed. Against this background, ε' along axis *a* begins to grow at temperatures *T* > 150 K, and there are two jumps of this growth at temperatures 150 and 250 K (Fig. 1a). Along axes *b* and *c*, only a growth of ε' with temperature begins from *T* ~ 225 K is observed (Figs. 1c and 1e, respectively). The data on the conductivities along axes *a*, *b*, *c* in the temperature range 80‑330 K shown in Figs. 1b, 1d, and 1f demonstrates that there are two temperature ranges, in which the character of the frequency dispersion of the conductivity is changed. At low temperatures up to some temperatures, which are dependent on the crystal axis orientation, the conductivity decreases as the frequency increases; i.e., at these temperatures, we have $\sigma_{DC} > \sigma_{AC}$. As temperature increases, the values of $\sigma_{AC}$ increase and, at some temperatures, at various frequencies, intersect the low temperature conductivity that is independent on temperature and does not have a frequency dispersion. The higher are the frequencies, at the higher temperatures, the more likely these intersections are observed. The frequency-dependent temperatures, at which $\sigma_{AC}$ are equalized with $\sigma_{DC}$, obey the Arrhenius law $\omega = \omega_0 \exp(-E_A/kT_m)$, where ω is frequency, $T_m$ is temperature at which $\sigma_{AC} = \sigma_{DC}$, $E_A$ is the activation energy at the boundary of local domains. This fact gives the possibility to determine the activation barriers at the boundaries of such low-temperature local domains for various crystal axes (the insets in Figs. 1b, 1d, and 1f). As is seen from these figures, along all the axes, there are additional high-temperature maxima $\sigma_{loc}$, the shifts of whose temperatures in the dependence on frequency also obey the Arrhenius law, which allows us to determine activation barriers at the boundaries of these high-temperature local domains. They are also shown in the plots for local conductivities. Thus, the observation of the local conductivity along all main axes of ErMn2O5 with different activation barriers demonstrates the existence of the phase separation domains of two types.

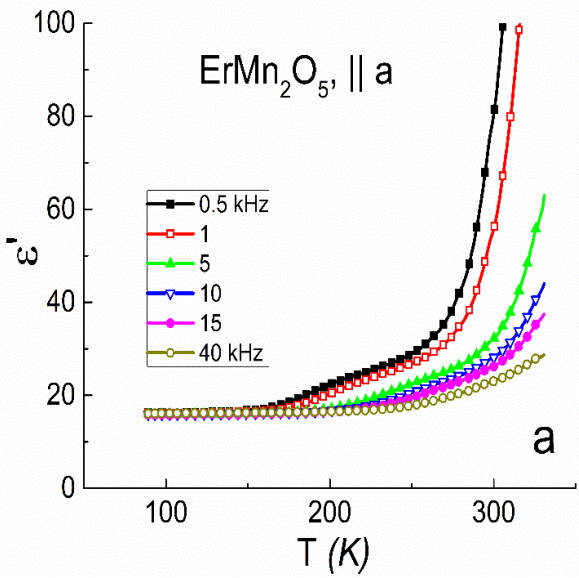
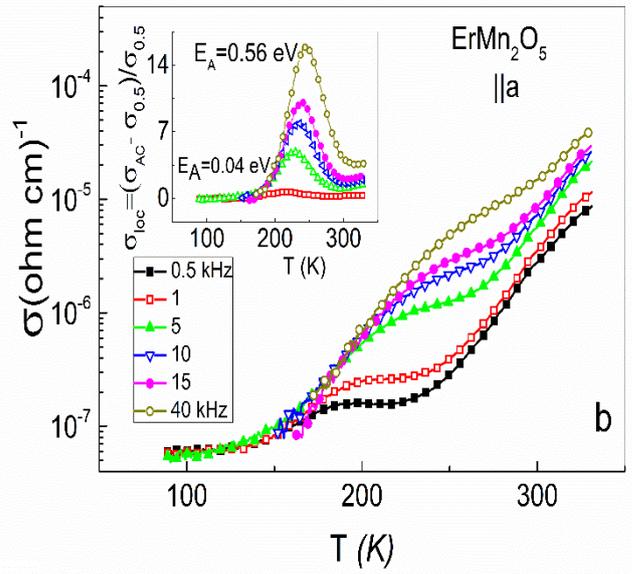
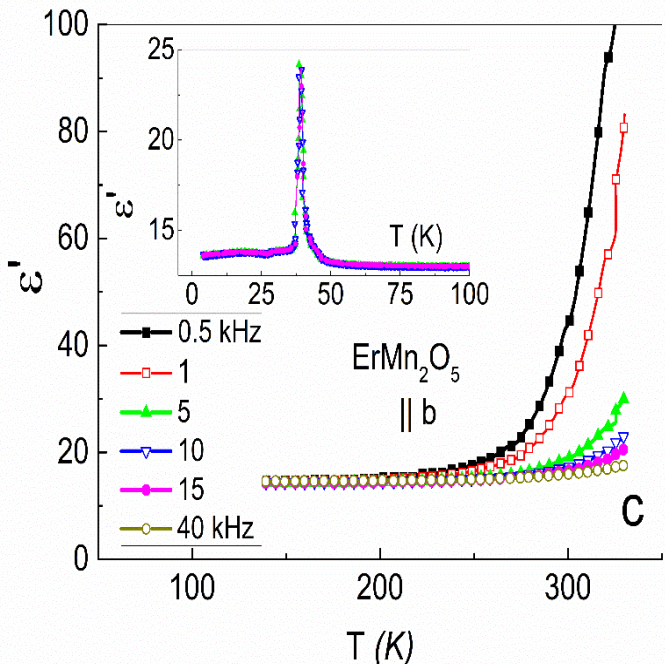
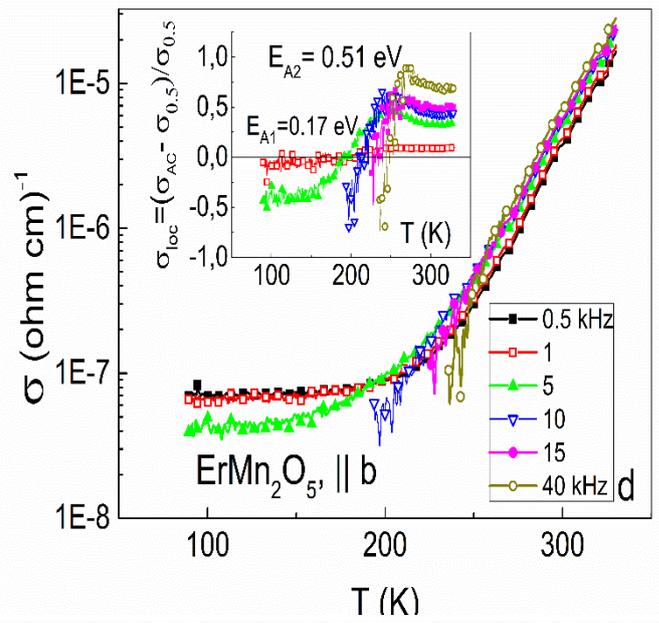
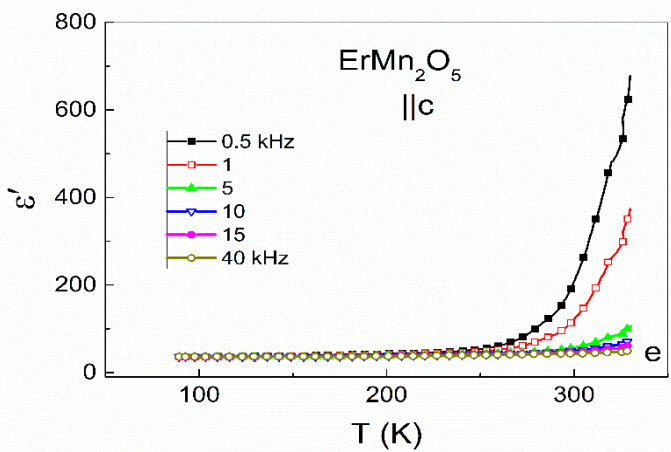
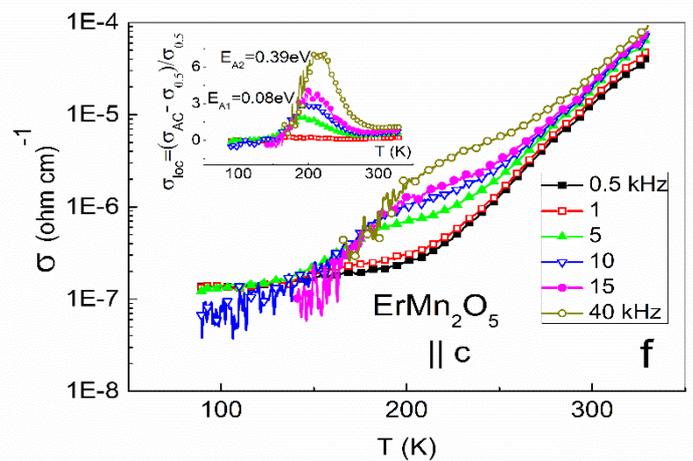

Fig. 1. Temperature dependences of permittivity ε' for a number of frequencies along axes *a*, *b*, and *c* (a, c, e, respectively) and conductivity σ along these axes (b, d, and f, respectively) for ErMn2O5. The insets in (b, d, and f) show the temperature dependences of the local conductivity. The frequencies are indicated in the plots.

We consider the regions with negative local conductivity $\sigma_{loc}$ (i.e., with low-frequency percolation dispersion free conductivity $\sigma_{DC}$ that is higher than high frequency conductivities $\sigma_{AC}$) as local phase separation domains, which form inside the manganese subsystem that is at low temperatures under high lattice barriers due to ions $Er^{3+}$ with the strong spin–orbit coupling. As is shown in [5, 9, 14–16] for RMn$_2$O$_5$ (R = Eu, Gd, and Bi) at low temperatures, the phase separation domains occupy a small crystal volume. At temperatures increased with the increase in the frequency, the probability for charge carriers to overcome high lattice barriers due to $Er^{3+}$ ions increase with temperature. In this case, high-temperature local phase separation domains form (Figs. 1b, 1d, and 1f). Note that the conductivity in these domains also is determined by transfers of $e_g$ electrons between manganese ions of various valences, but already in the domains with higher barriers. Thus, the local phase separation domains of the Mn subsystem exhibit different properties at quite low temperatures and at higher temperatures after overcoming a lattice barrier. The main characteristics of these domains are the temperature dependences of local conductivities $\sigma_{loc}$ (the insets in Figs. 1b, 1d, and 1f) and the electric polarization (Figs. 2a and 2b).

In [7–9], it was shown for GdMn$_2$O$_5$ and BiMn$_2$O$_5$ that the frozen superparamagnetic state, for which the electric polarization loops were observed, exists to a temperature, at which condition $\sigma_{loc} \approx 0$ obeys. Condition $\sigma_{loc} \approx 0$ is fulfilled at temperatures, at which $kT \approx E_A$. At higher temperatures, the phase-separation domains of the Mn subsystem continue to exist being in the conventional superparaelectric state, for which the electric polarization disappears.

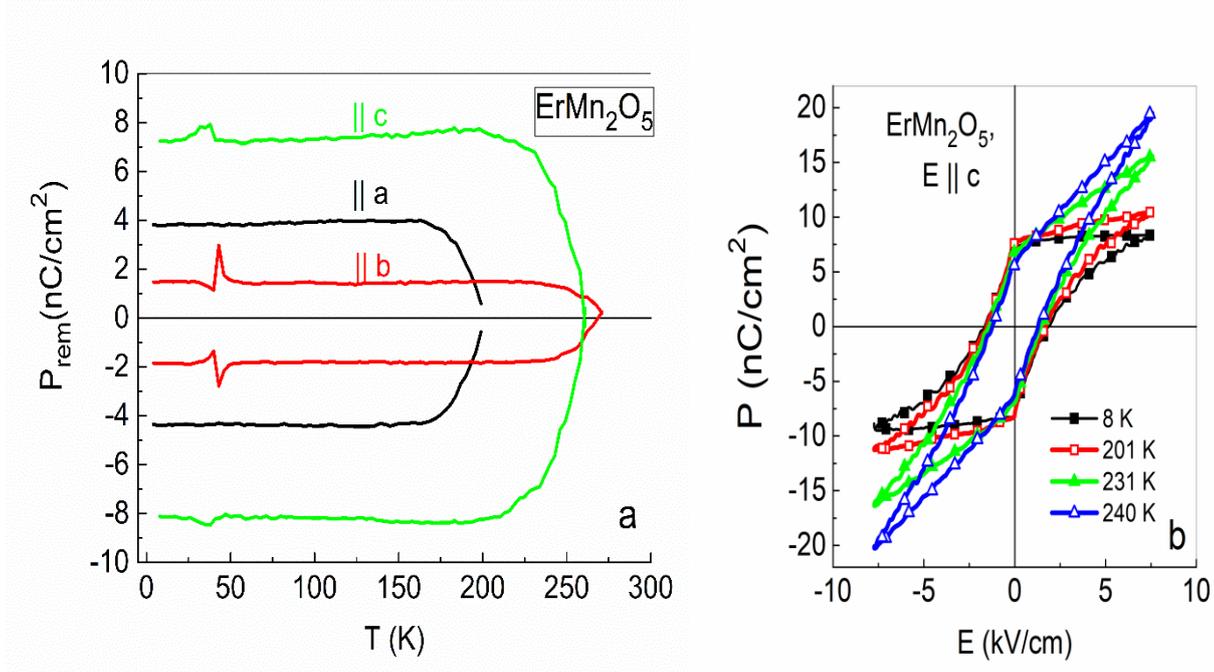

**Fig. 2.** Temperature dependences of remanent polarization $P_{rem}$ along various axes of $ErMn_2O_5$ (a) and a set of the electric polarization hysteresis loops at various temperatures along axis $c$ of $ErMn_2O_5$ (b).

In $ErMn_2O_5$, along axis $a$, near 190–200 K (Fig. 1b), the percolation conductivity (the conductivity at the lowest frequency) of the matrix increases stepwise. It would appear reasonable that this increase is due to thermal activation of charge carriers from the low temperature phase separation domains as condition $kT \approx E_A \approx 0.04$ eV is fulfilled for activation barriers of these domains. It is seen from Fig. 1b that, at the same temperatures, an increased percolation conductivity of the matrix approaches in magnitude the values of conductivities $\sigma_{AC}$ coinciding to each other. In this case, condition $\sigma_{loc} \approx 0$ is fulfilled at all frequencies, at which local polar domains change their state from the frozen superparaelectric to the convention superparaelectric state, and the electric polarization of phase separation domains sharply decrease to a zero (Fig. 2a). In $ErMn_2O_5$, the activation barrier at the low temperature phase separation domain boundaries along axis $b$ is significantly higher (0.18 eV) and approaching the values of the increasing percolation conductivity of the matrix and the coinciding $\sigma_{AC}$ of high-temperature phase separation domains occurs near 270 K (Fig. 1d). In this case, over the entire temperature range to 270 K, the values of $\sigma_{loc}$ are small (the inset in Fig. 1d), which decreases the value of the polarization observed along axis $b$ (Fig. 2a). Along axis $c$, at a temperature of 175 K, conductivities $\sigma_{AC}$ of high-temperature phase separation domains coinciding in magnitude are strongly different from the percolation conductivity of the matrix (Fig. 1f). The convergence of these values begins from 200 K, and, at a temperature of 260 K, the values of $\sigma_{loc}$ and the polarization approach zero (the inset in Figs. 1f and 2a).

Now we consider the electric polarization hysteresis loops in $ErMn_2O_5$ induced by the local polar phase separation domains. Figure 2a shows the temperature dependences of the remanent polarization along various axes of $ErMn_2O_5$. It is seen that the maximum residual polarization is observed along axis $c$ up to a temperature of 260 K. The minimum polarization appears along axis $b$ and exists up to a temperature of 270 K. Along axis $a$, the polarization observed to $T < 190$ K has an intermediate value. Figure 2b shows a set of the hysteresis loops for some temperatures along axis $c$. It is seen that the shape and the value of are slightly changed in the temperature range 8–200 K and quite sharply decrease in the temperature range 200–260 K. Note that the hysteresis loop shapes along all the axes are similar, and Fig. 2a shows the values of the polarization and the temperature of their existence.

The comparison of the data of Figs. 2a and 1 for the permittivity and the conductivities allow us to make the conclusion on the nature of the polarizations observed along various axes of $ErMn_2O_5$. Actually, it is precisely local polar phase separation domains that form the electric polarization observed in this work. The choice of the high-frequency regime of the PUND method of measuring the hysteresis loops and also the temperature range of their existence suggest that the response on applied electric field pulses is due to the high-dynamic domains. They are local polar phase separation domains and the fluctuation region near the phase transition to the ferroelectric ordering [7-9].

There is a clear correlation between the dielectric properties determined by the phase separation domains and the parameters of the observed electric polarization. The comparison of Figs. 1 and 2 for the corresponding directions of the crystal shows that there is the optimal relation of the activation barriers of the phase separation domains of the Mn-subsystem under the barriers due to $Er^{3+}$ ions and after overcoming these barriers. There is, as well, the optimal value of the local conductivity for the maximum polarization that exists to high temperatures. As noted above, the maximum polarization ($\approx$7.5 nC/cm$^2$) is observed up to a temperature of 260 K along axis *c*. In this case, the ratio of the activation barriers of the two types of domains (0.39/0.08 $\approx$ 5), the value $\sigma_{loc} \approx 7$ for the high-temperature maximum. These parameters can be considered as optimal for $ErMn_2O_5$. The minimum polarization ($\approx$2 nC/cm$^2$) is observed up to a temperature of 270 K along axis *b* (Fig. 2a). In this case, the ratio of the activation barriers of the two types of domains (0.51/0.18 $\approx$ 2.8) and the value $\sigma_{loc} < 1$. Along axis *a*, the ratio of the activation barriers of the two types of domains (0.56/0.04 $\approx$ 14), the value $\sigma_{loc}$ is $\approx$17 for the high-temperature maximum. In this case, the polarization ($\approx$4 nC/cm$^2$) is observed to relatively low temperature of 190 K. As a result, we conclude that the proximity of the activation barrier of the two types of the local domains substantially decreases the local conductivity and polarization, but provides the existence of the polarization to higher temperature (along axis *b*). The very large difference of the activation barriers along axis *a* leads to a sharp increase in $\sigma_{loc}$ for the high-temperature maximum. As a result, within such local domains, excess (not bounded by the charge changing of $Mn^{3+}$–$Mn^{4+}$ ion pairs) carriers appears under the barrier, and they screen the barrier and decrease the polarization, as well as decreases also the temperature of existence of the polarization (axis *a*). Note that, along axis *a*, additional lower-temperature jumps appear in the temperature dependences of permittivity $\varepsilon'$; the jumps are dependent on frequency and are thought to be related with the appearance of a comparatively low-temperature frozen superparaelectric state that give the contribution to the polarization of the phase separation domains. The higher-temperature domains of the local conductivity refer to the conventional superparaelectric state that does not give the contribution to the electric polarization loops. As a result, the temperatures of disappearance of the electric polarization in the temperature dependences of the permittivity (Figs. 1a, 1c, and 1e) coincide with the lowest-temperature increases in values of $\varepsilon'$, which are dependent on the frequency.

### 3.2. Dielectric Properties and the Electric Polarization in TbMn2O5

Taking into account the analysis of the correlation of the data on the dielectric properties and the polarization for $ErMn_2O_5$ presented above, we present, firstly, the data on the electric polarization for $TbMn_2O_5$ and then will analyze their dielectric properties for each of the crystal axes. In this case, we will compare the situations in $TbMn_2O_5$ and $ErMn_2O_5$ along the corresponding axes. Figures 3a and 3b show the temperature dependence of the remanent polarization along different axes of $TbMn_2O_5$ and the hysteresis loops of the electric polarization along axis *a*. It is seen that, unlike $ErMn_2O_5$ (Fig. 2a), the values of the remanent polarization along all axes in $TbMn_2O_5$ are close to each other and they are lower than those in $ErMn_2O_5$ along axes *a* and, particularly, along axis *c*. The values of the remanent polarizations in two of these crystals are close to each other along axis *b*. We relate this fact to a significantly higher value of the orbital moment of $Er^{3+}$ ions strongly coupled (as compared to $Tb^{3+}$ ions) with the lattice.

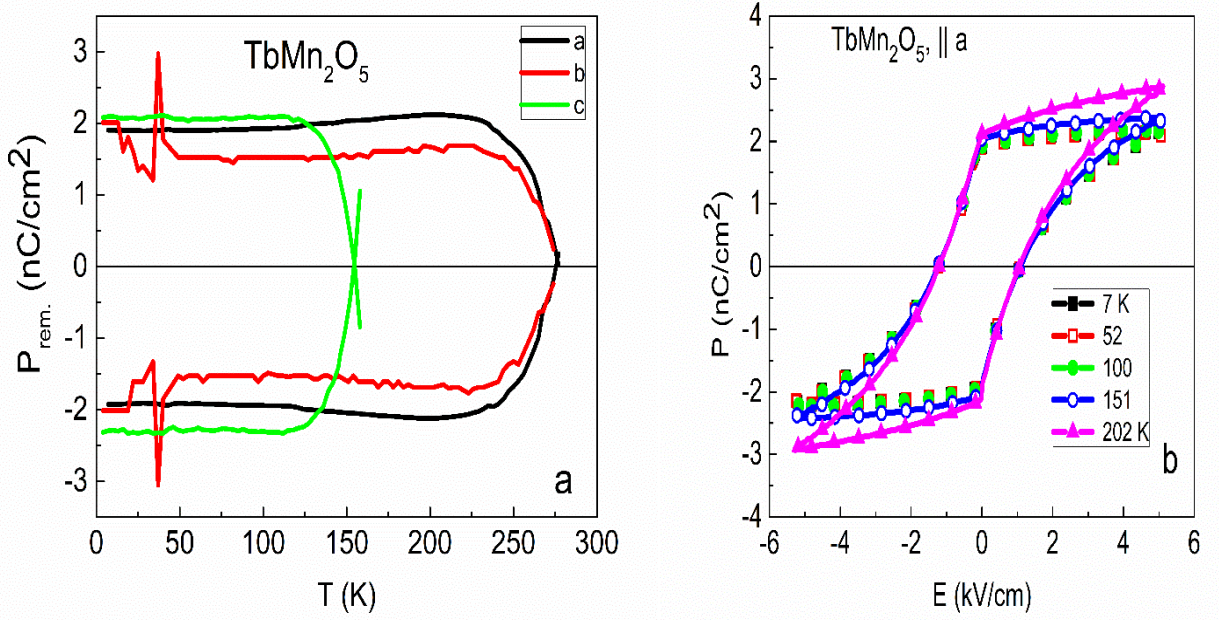

**Fig. 3.** Temperature dependences of remanent polarization $P$rem along various axes of TbMn$_2$O$_5$ (a) and a set of the electric polarization hysteresis loops at various temperatures along axis *a* of TbMn$_2$O$_5$ (b).

In this case, as noted above, the moment in ErMn$_2$O$_5$ is mainly directed along axis *c*. In TbMn$_2$O$_5$, the orbital moment is oriented in plane *ab*. Figures 4a, 4c, and 4e show the temperature dependences of the permittivity $\varepsilon'$ for a number of frequencies in TbMn$_2$O$_5$ along axes *a*, *b*, and *c*, respectively. Figures 4b, 4d, and 4f demonstrate the temperature dependences of the conductivities for the same frequencies along the same axes. The inset in these figures show the corresponding temperature dependences of $\sigma_{loc} = (\sigma_{AC} - \sigma_{DC})/\sigma_{DC}$. In TbMn$_2$O$_5$ as is in ErMn$_2$O$_5$, there exist both low-temperature and high temperature local phase separation domains along the all axes.

Along axis *a*, the conductivity in TbMn$_2$O$_5$ in the temperature range 5–275 K (Fig. 4b) is an order of value lower than that in ErMn$_2$O$_5$ (Fig. 1b). This fact determines a qualitative change in all parameters in TbMn$_2$O$_5$ as compared to ErMn$_2$O$_5$. Actually, the value of $\varepsilon'$ in TbMn$_2$O$_5$ along axis *a* (Fig. 4a) is significantly lower than that in ErMn$_2$O$_5$ (Fig. 1a).

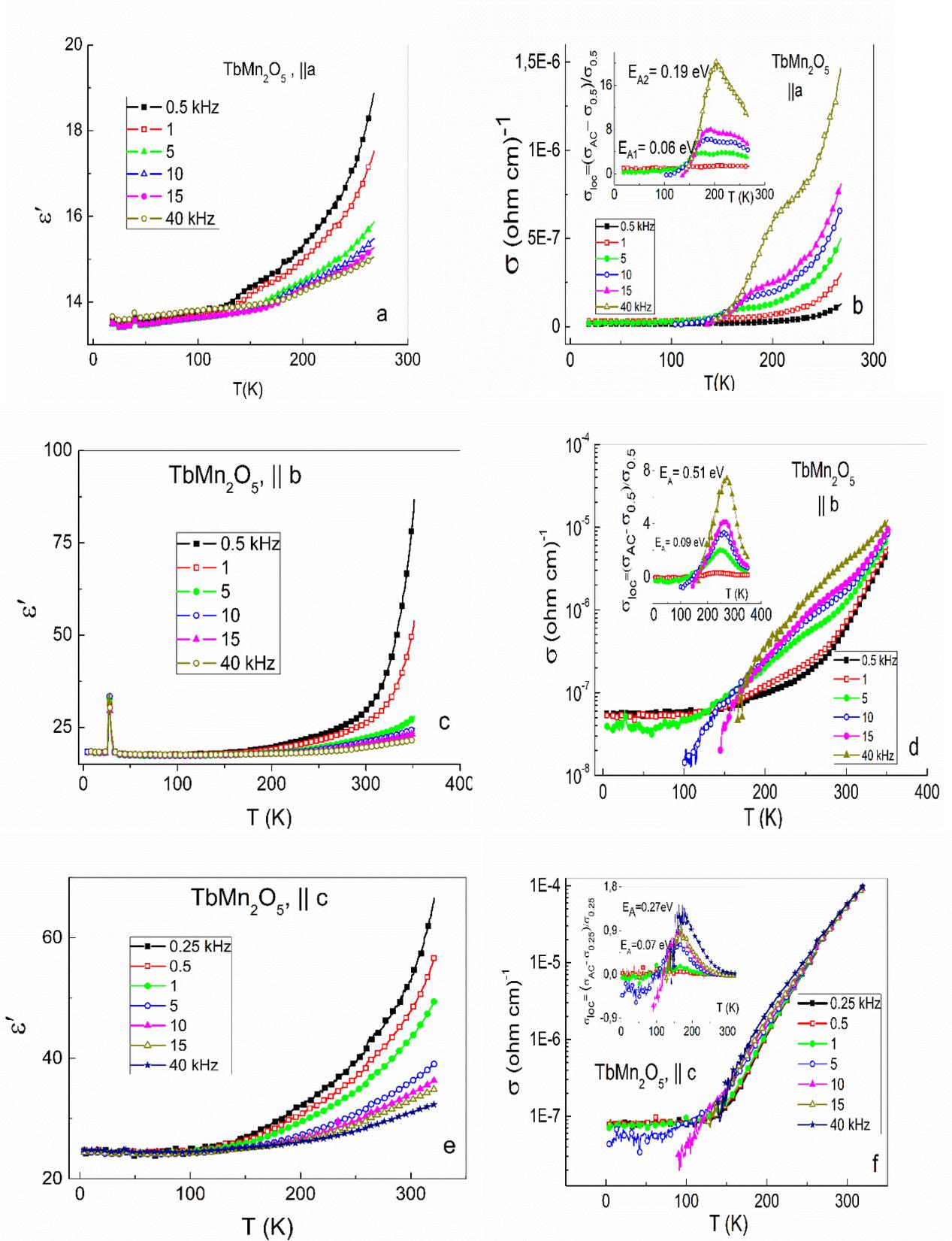

**Fig. 4.** Temperature dependences of permittivity ε' for a number of frequencies along axes *a*, *b*, and *c* (Figs. 4a, 4c, and 4e, respectively) and conductivity σ along these axes (Figs. 4b, 4d, and 4f, respectively) for TbMn$_2$O$_5$. The insets in Figs. 4b, 4d, and 4f show the temperature dependences of the local conductivity. The frequencies are indicated in the plots.

As a result, the first jump of ε' in TbMn$_2$O$_5$ near 120 K only indicates the region of insignificant increase in polarization that continues to exist to higher temperature 275 K (Fig. 3a). In TbMn$_2$O$_5$ as is in ErMn$_2$O$_5$, along axis *a*, the barriers at the boundaries of the low temperature domains are commensurate (0.06 and 0.04 eV, respectively), while the barriers at the boundaries of high-temperature domains in ErMn$_2$O$_5$ (0.56 eV) (Fig. 1b) are higher than those in TbMn$_2$O$_5$ (0.19 eV) by a factor of almost three (Fig. 4b). The high-temperature local domains in TbMn$_2$O$_5$ (as in ErMn$_2$O$_5$) exhibit close high relative local conductivities ($\sigma_{loc} \approx 18$) against the background of the one order- smaller values of $\sigma_{AC}$ and $\sigma_{DC}$ (in TbMn$_2$O$_5$, they are lower). As a result, in TbMn$_2$O5, the internal field of the high-temperature domains is not completely screened up to 275 K, but at significantly weakened polarization (Fig. 3a). Recall that, in ErMn$_2$O$_5$, the screening of the internal field of the high-temperature domains was observed near 200 K, but at P$_{rem}$ that was higher by a factor of two (Fig. 2a).

The conductivities of both crystals along axis *b* are slightly different, and this is the determining factor of the similar properties of the electric polarization in ErMn$_2$O$_5$ and TbMn$_2$O$_5$ (compare Figs. 2a and 3a) along this axis. We relate this circumstance to the fact that, in all RMn$_2$O$_5$, independent of the type of R ions, the Mn$^{3+}$–Mn$^{4+}$ ion pairs alternate along axis *b*, and the increased conductivity along this axis is due to the transfer of *e$_g$* electrons between these ion pairs. ErMn$_2$O$_5$ and TbMn$_2$O$_5$ demonstrate close values P$_{rem}$ and the temperatures of existence of the electric polarization along the *b* axis.

The conductivities along axis *c* in TbMn$_2$O$_5$ and ErMn$_2$O$_5$ (Figs. 4f and 2f, respectively) are also close to one another, which also is related to the fact that the formation of the phase separation domains in all RMn$_2$O$_5$ at all temperatures is due to the transfer of *e$_g$* electrons between the neighboring planes which are perpendicular to axis *c* and contain only Mn$^{3+}$ or Mn$^{4+}$ ions. However, the permittivities and local conductivities in TbMn$_2$O$_5$ are substantially lower than those in ErMn$_2$O$_5$ (compare Figs. 4e and 1e and also Figs. 4f and 1f, respectively), which leads to a qualitative difference of the values and the temperatures of existence of the electric polarizations. In ErMn$_2$O$_5$, the polarization of ~7.5 nC/cm$^2$ along axis $c$ exists up to a temperature of 260 K, while in TbMn$_2$O$_5$, the polarization of ~2 nC/cm$^2$ along axis $c$ exists up to a temperature of 154 K (Figs. 2a and 3a, respectively). This is explained by the fact that Er$^{3+}$ ions strongly distort the lattice precisely along axis $c$, while Tb$^{3+}$ ions influence in the $ab$ plane. As it was the case in ErMn$_2$O$_5$, the temperature increase in the permittivity of TbMn$_2$O$_5$ agrees with the temperatures of disappearance of the electric polarization (Fig. 3a).

Attention is drawn to the fact that the lowest and quite close temperatures of existence of the electric polarization are observed in ErMn$_2$O$_5$ along axis $a$ (Fig. 2b) and in TbMn$_2$O$_5$ along axis $c$ (Fig. 4f). In both cases, they are determined by the phase transformation temperatures, at which all high-frequency conductivity of the low-temperature domains are compared to one another against the background of weak local conductivities of the high-temperature domains (Figs. 1b and 4f, respectively). In this case, the high-temperature domains form the conventional superparaelectric state, for which electric polarization does not exist. This fact is also due to the circumstance that Er$^{3+}$ and Tb$^{3+}$ ions do not distort the lattices exactly along axes $a$ and $c$, respectively. This is why, in TbMn$_2$O$_5$, the weakly-dispersion conductivity sharply increases at $T > 150$ K along axis $c$ with a small contribution of the local conductivity (Fig. 4f). As noted above, in ErMn$_2$O$_5$ along axis $a$ (Fig. 1b), at $T > 190$ K, in contrast, the high-temperature local domains with anomalously high local conductivity appear inside these domains, which lead to the screening of the internal electric fields and the electric polarization.

Note that the substantial difference in the behavior of the local conductivities and the electric polarization induced by local polar phase separation domains in GdMn$_2$O$_5$ [9–11] and RMn$_2$O$_5$ (R = Er, Tb) is due to the absence in GdMn$_2$O$_5$ of additional barriers in the lattice, which are formed by Er$^{3+}$ and Tb$^{3+}$ ions with strong spin-orbit interactions. In the ground state of Gd$^{3+}$ ions, there is no spin-orbit interaction. As a result, in GdMn$_2$O$_5$, the local domains of both the types with different barriers at their boundaries do not coexist along the same axes. As indicated above, it is precisely the interaction of these two types of domains with crystal fields of different values, that determine the temperature evolution of the dielectric properties and the polarization in RMn$_2$O$_5$ (R = Er, Tb).

## 4. CONCLUSIONS

It is shown that various R ions in RMn$_2$O$_5$ qualitatively change the properties of the local phase separation domains and the electric polarization induced by them. The appearance of phase separation domains is determined by the states of the Mn subsystem (charge ordering of ions of different valences (Mn$^{3+}$ and Mn$^{4+}$) and their charge exchange by $e_g$ electrons). However, these domains are in different crystal fields of the lattice depending on the type of *R* ions. Gd$^{3+}$ ions in GdMn$_2$O$_5$ do not distort the lattice, and the local polar domains and the electric polarization observed in them reflect the properties of the Mn subsystem itself. Conversely, ions Er$^{3+}$ and Tb$^{3+}$ in RMn$_2$O$_5$ are strongly and variously distort the crystal fields, in which there are Mn$^{3+}$ and Mn$^{4+}$ ions and electrons, which recharge them. As a result, local domains of two the types appear in the Mn subsystem. Below certain temperatures, the local phase separation domains of the Mn subsystem are under the barrier in the crystal field that is formed by *R* ions. As temperature increases, charge carriers begin to overcome these barriers in the lattice, and the phase separation domains and the electric polarization induced by them change their properties. Since the spin–orbit interactions for Er$^{3+}$ and Tb$^{3+}$ ions differ in the value and the anisotropy, the dielectric properties and the polarization in ErMn$_2$O$_5$ and $TbMn_2O_5$ are strongly different form each other, and both these crystals have different properties as compared to $GdMn_2O_5$ [7–9].

FUNDING

This work was supported by the Russian Foundation for Basic Research (project nos. 18-32-00241) and Program 1.4 of the Presidium of the Russian Academy of Sciences "Actual problems of the low-temperature physics."

CONFLICT OF INTEREST

The authors declare that they have no conflicts of interest.